\documentclass[prl,aps,twocolumn,showpacs,superscriptaddress,10pt]{revtex4-1}
\usepackage{amsmath,amssymb,bm,graphicx,dsfont,color}
\usepackage[latin9]{inputenc}
\usepackage{amsmath}
\usepackage{slashed}
\usepackage{dsfont}
\usepackage{amstext}
\usepackage{amssymb}
\usepackage{amsbsy}
\usepackage{amsthm}
\usepackage{epsfig}
\usepackage{graphicx}
\usepackage{bbm}
\usepackage{hyperref}
\usepackage{color}
 
\begin{document}


\title{Microscopic model for the boson integer quantum Hall effect }
\author{N. Regnault}
\affiliation{Department of Physics, Princeton University, Princeton, NJ 08544}
\affiliation{Laboratoire Pierre Aigrain, ENS and CNRS, 24 rue Lhomond, 75005 Paris, France}
\author{T. Senthil}
\affiliation{Department of Physics, Massachusetts Institute of Technology,
Cambridge, MA 02139, USA}
 \date{\today}
\begin{abstract}
In two dimensions strongly interacting bosons in a magnetic field can form an integer quantum Hall state. This state has a bulk gap, no fractional charges or topological order in the bulk but nevertheless has quantized Hall transport and symmetry protected edge excitations. Here we study a simple microscopic model for such a state in a system of two component bosons in a strong orbital magnetic field. We show through exact diagonalization calculations that the model supports the boson integer quantum Hall ground state in a range of parameters. 

\end{abstract}
\newcommand{\be}{\begin{equation}}
\newcommand{\ee}{\end{equation}}
\maketitle

Recently there has been considerable theoretical activity on the possibility of ``Symmetry Protected Topological (SPT)" insulating phases of interacting bosons in diverse dimensions. 
These are generalizations of the celebrated free fermion topological insulators\cite{TIs} to systems of bosons. As in the non-interacting limit a boson system is trivial, strong interactions are necessary to reach any insulating phase, including the topological ones. Thus studies  of bosonic insulators force us to think of topological insulation without the crutch of free fermions and band topology appropriate for non-interaction fermion systems. It is expected that the insights from these studies will be useful in thinking about the interplay of strong correlations with topological insulation in electronic systems. 

In the interacting context we seek phases of bosonic matter that have a bulk gap, have no `intrinsic' topological order or fractionalized excitations, but nevertheless have protected boundary excitations. 
In space dimension $d = 1$ the familiar Haldane spin chain provides a well established example of such a state. Recently progress was made in classifying all such gapped $1d$ systems based on the concept of group cohomology\cite{Fidkowski2011,Turner2011,Chen2011a,Cirac}. In $d > 1$ a cohomology classification has also been proposed\cite{chencoho2011} of all gapped bosonic systems with no `intrinsic' topological order which predicts the existence of many SPT states. Physical properties of these phases are not readily accessed through this framework and have been addressed through other methods in both two\cite{LevinGu,luav2012,tsml13} and three dimensions\cite{avts12,xuts13,hmodl,maxetal13}.  States beyond the cohomology classification are also known\cite{avts12,hmodl,burnell13} to exist (in $d = 3$).

One of the most important questions that remains open in this field is the identification of experimentally relevant microscopic models for these SPT phases in $d > 1$.  
Here we address this question for a prototypical SPT phase in $d = 2$, namely an integer quantum Hall state of bosons. We show that a simple and realistic Hamiltonian for a 
system of two-component bosons in a strong magnetic field provides a concrete realization of this state. The physical properties of this state have been studied in Refs.~\onlinecite{luav2012,tsml13,liuxgw,lesik13}.  The electrical  Hall conductivity $\sigma_{xy}$ was shown\cite{luav2012} to be quantized to be an even integer (see Ref.~\onlinecite{tsml13} for a simple argument).  At the edge to the vacuum there is a charged chiral edge mode and a counter propagating neutral mode. Despite being non-chiral this edge structure is protected so long as charge conservation symmetry is preserved. This boson integer quantum Hall state provides a prototypical example of a Symmetry Protected Topological insulator in boson systems in two space dimensions. Ref.~\onlinecite{tsml13} proposed that two component bosons in strong orbital magnetic fields at filling factor $\nu = 1$ for each component provide a suitable platform to realize the boson integer quantum Hall state (bIQH). A number of potential competing states were also identified. 

In this paper we study the simplest Hamiltonian for such a system through exact diagonalization studies. We find strong evidence for the stability of the boson integer quantum Hall state. 

\begin{figure}[htb]
\includegraphics[width=0.94\linewidth]{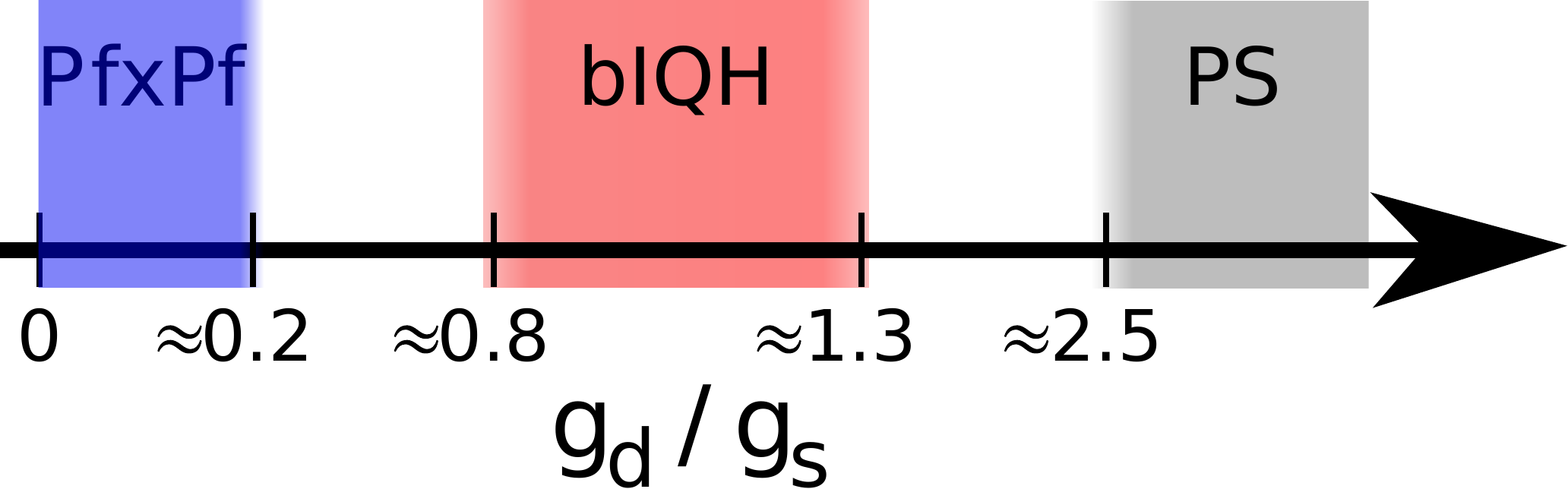}
\caption{Phase diagram of the model Hamiltonian. The bIQH state is stable in a range of $\frac{g_d}{g_s}$ estimated to be between $\sim 0.8$ and $\sim 1.3$ from exact diagonalization calculations with 12 and 16 particles. PS refers to the phase separated state which appears for large $\frac{g_d}{g_s}$. For small $\frac{g_d}{g_s}$ the decoupled Pfaffian state ${\mathrm{Pf}}\times {\mathrm{Pf}}$ is stable.}
\label{fig:biqhpd}
\end{figure}

We consider a bilayer system of $N$ interacting bosons and $N_\Phi$ flux quanta. All our calculations are performed in the lowest Landau level and the filling factor is defined as $\nu_t =N / N_{\Phi}$. The bilayer index is equivalent to a spin $1/2$. The number of particles per layer is respectively $N^\uparrow$ and $N^\downarrow$. We define the spin projection as $S_z=(N^\uparrow - N^\downarrow) / 2$. We denote by ${z^\uparrow_i,z^\downarrow_i}$ the complex coordinates of the two boson species. The Hamiltonian projected in the lowest Landau level reads:
\begin{eqnarray}
{\cal H}&=&g_s \sum_{i<j} \delta^{(2)} \left(z^\uparrow_{i}-z^\uparrow_{j}\right) + g_s \delta^{(2)} \left(z^\downarrow_{i}-z^\downarrow_{j}\right)\nonumber\\
& & + g_d \sum_{i,j} \delta^{(2)} \left(z^\uparrow_{i}-z^\downarrow_{j}\right)\label{eq:hamiltonian}
\end{eqnarray}
 When $g_s = g_d$ the Hamiltonian has an extra pseudospin $SU(2)$ symmetry which rotates the two species of bosons into one another.  The ground state of this model  was previously studied using exact diagonalization in Refs.~\onlinecite{gjbl12,ueda12}  with an emphasis on total filling factor $\nu_t = \frac{4}{3}$.  Here we focus instead on $\nu_t = 2$ which is a suitable platform to realize the bIQH state.  Ref.~\onlinecite{tsml13} showed that prototypical wave functions for the bIQH state is a pseudo spin singlet. Thus we may expect the bIQH state to be stabilized, if at all, near the $g_s = g_d$ point. A competing pseudo spin singlet state is a member of a family of Non-Abelian Spin Singlet (NASS) states\cite{ardonne99} proposed for two-component bosons at various total fillings $\nu_t = \frac{2k}{3}$ ($k$ integer).  Our calculations below show that the bIQH state wins over the NASS state at total filling factor $\nu_t = 1+1 = 2$.  When the ratio $\frac{g_d}{g_s} = 0$, the two components are decoupled and at $\nu = 1$ each. With delta function repulsion there is strong evidence\cite{Cooper-PhysRevLett.87.120405,Regnault-PhysRevLett.91.030402,Chang-PhysRevA.72.013611} that each component is in the non-Abelian Pfaffian (Moore-Read) state (Pf)\cite{Moore:1991p165}. This decoupled ${\rm Pf}\times {\rm Pf}$ state of the 2-component system will be stable in some range of small $\frac{g_d}{g_s}$. The corresponding wave function reads 
\begin{eqnarray}
\Psi_{{\rm Pf}\times {\rm Pf}} &=& {\rm Pf} \left(\frac{1}{z^\uparrow_i - z^\uparrow_j}\right) \prod_{i < j} (z^\uparrow_i - z^\uparrow_j)\nonumber \\
	&&\times {\rm Pf} \left(\frac{1}{z^\downarrow_i - z^\downarrow_j}\right) \prod_{i < j} (z^\downarrow_i - z^\downarrow_j) \label{wfPfPf}
\end{eqnarray}
In the other limit when $\frac{g_d}{g_s} = \infty$ we expect phase separation into puddles with each puddle consisting of bosons living in one of the two layers with twice the average density (hence at filling factor $\nu = 2$).  Thus we expect the region around $\frac{g_d}{g_s} = 1$ to be most favorable for the bIQH state and to see phase transitions to ${\rm Pf}\times {\rm Pf}$ or the phase separated state as this ratio is varied. Our calculations below confirm this expectation.  We will see that there is a sizable gap at the $SU(2)$ symmetric point which persists in a wide range of the ratio $\frac{g_d}{g_s}$. Thus the bIQH state seems quite robust in this model.  The phase diagram obtained from our work is summarized in Fig. ~\ref{fig:biqhpd}. 
 
 A wave function for the bIQH state was proposed in Ref.~\onlinecite{tsml13}: 
 \begin{eqnarray}
\Psi_{flux} = P_{LLL} \Bigl[&&\prod_{i < j} |z^\uparrow_i - z^\uparrow_j|^2 \cdot \prod_{i < j} |z^\downarrow_i - z^\downarrow_j|^2 \nonumber \\
&& \cdot\prod_{i , j} (z^\uparrow_i - z^\downarrow_j)\Bigr] \label{wf2}
\end{eqnarray}
where $P_{LLL}$ denotes the projection onto the lowest Landau level. As there is no intrinsic topological order there is a unique ground state on the torus. 
 
The competing NASS state can be built up starting with the $(m,m,n)$-Halperin wave function given by:    
\begin{equation}
 \Psi_{(m,m,n)} = \prod_{i< j} (z^\uparrow_{i}-z^\uparrow_{j})^{m} \prod_{i< j} (z^\downarrow_{i}-z^\downarrow_{j})^{m}\prod_{i,j} (z^\uparrow_{i}-z^\downarrow_{j})^{n} \label{halperin}
\end{equation}
The filling factor is $\nu_t =2/(m+n)$. Note that the $(1,1,1)$-Halperin is the densest zero energy state for our model Hamiltonian in Eq.~\ref{eq:hamiltonian} when $g_s, g_d > 0$. 
From the Halperin wave function, one can build a series of NASS states at filling $\nu_t=\frac{2k}{2 m - 1}$. This is done by dividing the particles into $k$ groups, writing a Halperin $(m;m,m-1)$ state for each group and then symmetrizing over the different groups. This procedure leads to:
\begin{eqnarray}
 \Psi_{m}^{k} = \mathcal{S}\Bigl[\prod_{i = 0}^{k-1}\Psi_{(m,m,m-1)}(&&z^\uparrow_{i\frac{N}{2k}+1},...,z^\uparrow_{(i+1)\frac{N}{2k}},\nonumber\\
	&& z^\downarrow_{i\frac{N}{2k}+1},...,z^\downarrow_{(i+1)\frac{N}{2k}})\Bigr]\label{eq:NASS}
\end{eqnarray}
where $\mathcal{S}$ is the symmetrization operator. For $m=2,n=1$, these are the NASS states introduced by Ref.~\cite{ardonne99}. The NASS state is $\frac{(k+2)(k+1)}{2}$ degenerate on the torus geometry.

\begin{figure}[htb]
\includegraphics[width=0.94\linewidth]{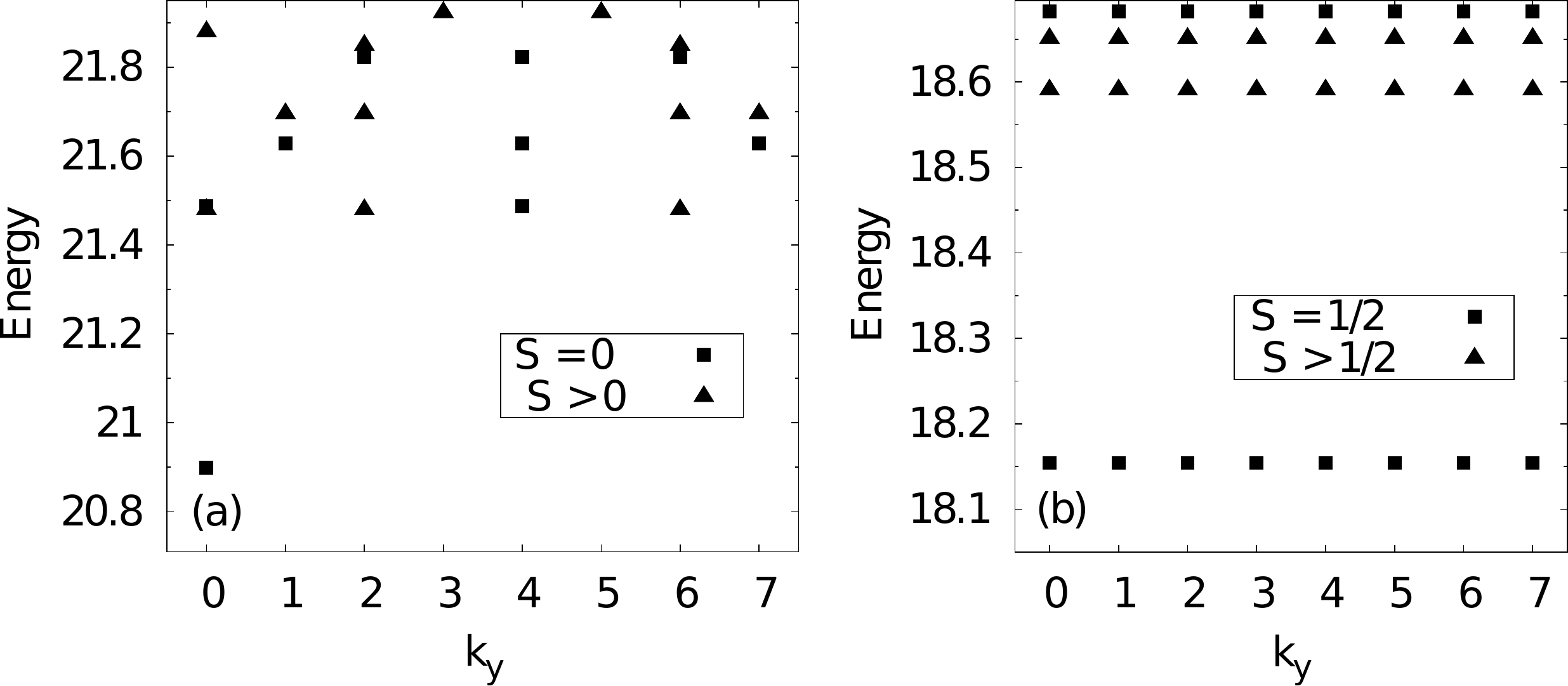}
\caption{{\it Left panel:} Low energy spectrum for $N=16$ bosons and $N_{\Phi}=8$ flux quanta on the torus geometry. The ground state is a spin singlet ($S=0$) clearly separated from the higher energy excitations. {\it Right panel:} Low energy spectrum for $N=15$ bosons and $N_{\Phi}=8$ flux quanta on the torus geometry. This situation corresponds to one less particle that the left panel. Since $N$ is odd, the spin is half integer. The exact degeneracy is a consequence of the magnetic translation symmetry when $N$ and $N_{\Phi}$ are relative prime numbers.}
\label{fig:spectrum}
\end{figure}

\begin{figure}[htb]
\includegraphics[width=0.8\linewidth]{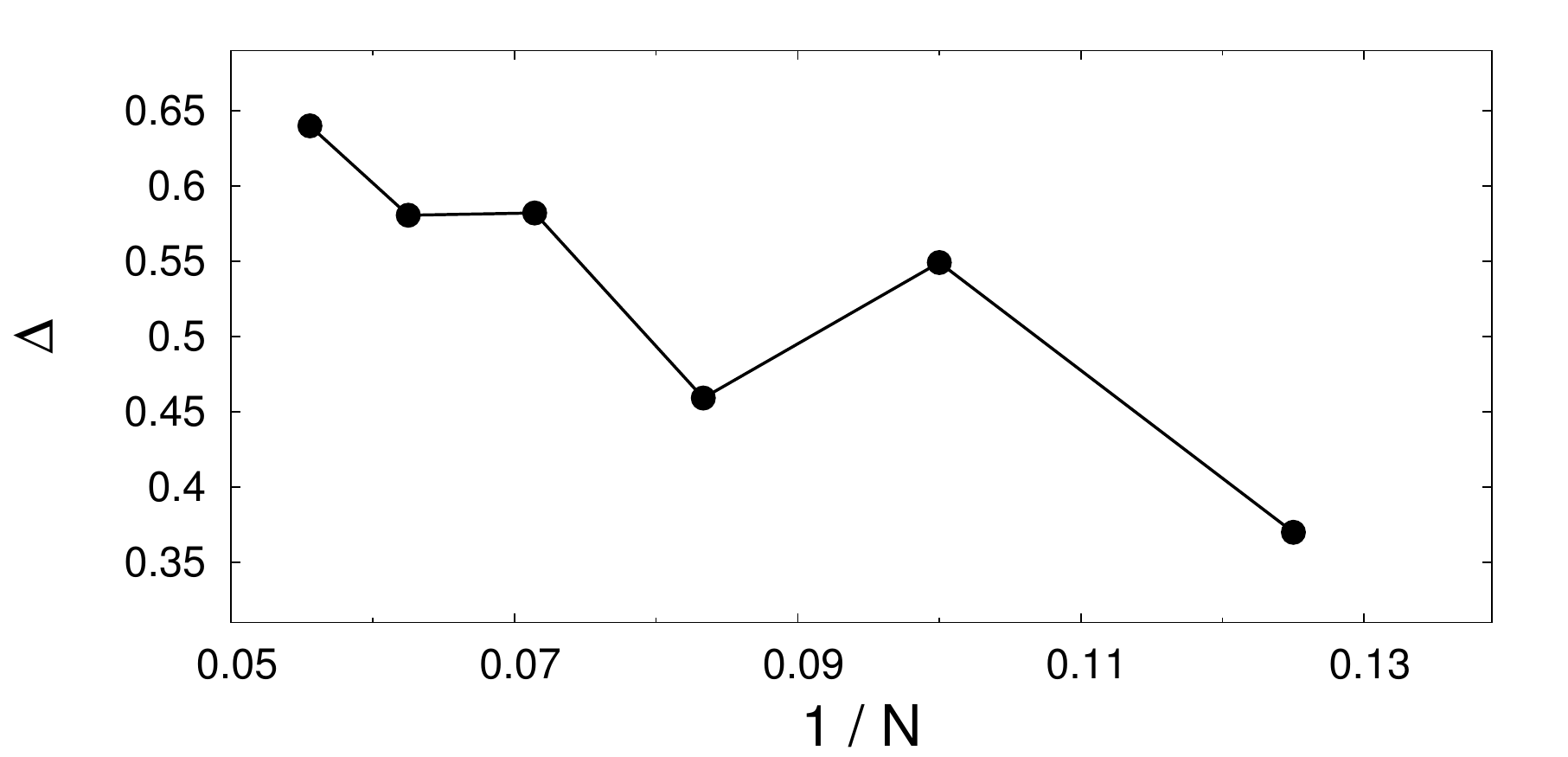}
\caption{Neutral gap $\Delta$ as a function of $1/N$ at filling factor $\nu=2$. While a thermodynamical extrapolation is quite difficult, the trend strongly suggests that the gap will stay finite for large $N$.}
\label{fig:gap}
\end{figure}

For our numerical calculations, we consider the torus geometry with a square aspect ratio. The choice of the torus instead of the sphere allows us to avoid the bias of the so-called shift: In many cases, such a bias prevents the direct comparison of possible candidates in finite size calculations. Moreover, each model state has a well defined non-trivial degeneracy on the torus, providing a simple signature that we can extract form the energy spectrum. We performed calculations using the translation symmetry of the torus along the $y$ direction. Thus our states are only labeled by the $k_y$ momentum which can be expressed as an integer between $0$ and $N_{\Phi}-1$. We first consider the $SU(2)$ symmetric point (i.e. $g_s=g_d$). We have looked at several system sizes up to $18$ particles. A typical energy spectrum is shown in Fig.~\ref{fig:spectrum}a for $N=16$ at filling factor $\nu=2$. We clearly observe a unique ground state at $k_y=0$ and with total spin $S=0$. Such a signature is the one expected for an IQH state at this filling factor. There is no trace left of the ${\mathrm{Pf}}\times {\mathrm{Pf}}$ that should occurs in the limit $g_d=0$. Its 9-fold degenerate manifold (5 states in momentum sector $k_y=0$ and 4 states in momentum sector $k_y=4$) is completely lifted. All the energy spectra at $\nu=2$ that we have computed share exactly the same features as this $N=16$ example. In Fig.~\ref{fig:gap}, we give the behavior of the neutral gap $\Delta$ (i.e. without changing the number of particles or the number of flux quanta) as a function of the system size. While a thermodynamical extrapolation is not straightforward due to the size effects, it looks convincing that $\Delta$ will stay finite. Note that contrary to most of the other FQH phase, the gap seems to slightly increase with increasing $N$.

The hallmark of the FQHE is the unique nature of its excitations which are realized by inserting or removing flux quanta or particles. In the case of the integer quantum Hall at $\nu=1$, removing on electron leads to a number of degenerate ground states that exactly matches the number of orbitals. For many of the model states of the FQHE, allows to predict how many states should appear in the low energy spectrum using the Haldane's exclusion principle\cite{haldane91}. Each counting is a signature of the considered phase. Fig.~\ref{fig:spectrum}b shows the low energy part of the spectrum when we remove one particle from the case described in Fig.~\ref{fig:spectrum}a. There is a set of low energy states with spin $S=\frac{1}{2}$, one per momentum sector. In this example, we thus get $16$ states. This is exactly what would be obtained for the bIQH state  when removing $1$ particle from a finite system $16$ fermions. Similar results are obtained when adding one particle. This is a clear indication that the low energy physics of our model is equivalent to the bIQH state. Other ways to nucleate excitations such as adding or removing flux quanta, generate much more states. Due to the energy splitting in finite size calculations, the signatures in the energy spectrum are not as clear as in the case of adding removing one particle.

We have tried to look at the particle entanglement spectrum\cite{li08,sterdyniak11} of the ground state. Unfortunately, we were not able to extract any clear signature. Note that the bIQH phase is  a composite fermion-like state with reverse flux attachment and such a behavior is expected: Such states (like the fermionic spin-polarized $2/3$ state) have a particle entanglement spectrum that looks a generic wave function for the same system size (i.e. the corresponding reduced density matrix does have neither an exponentially large number of zero eigenvalues nor an entanglement gap). We have also looked at the sphere geometry at the shift given by the model wave function of Eq.~\ref{wf2}. For every system size, we have observed a spin-single state with a total angular momentum equal to zero and a large gap.

\begin{figure}[htb]
\includegraphics[width=0.98\linewidth]{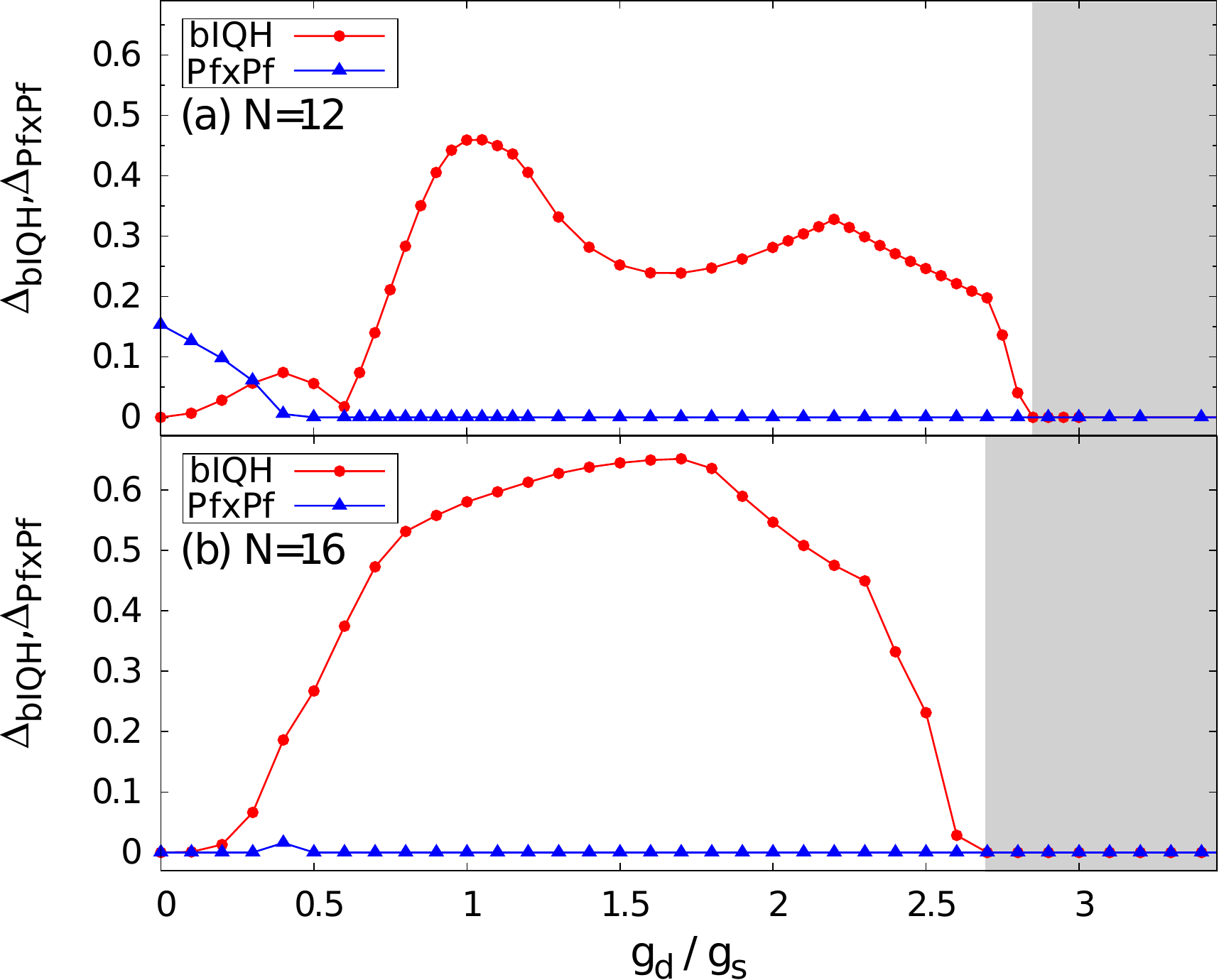}
\caption{Phase diagram for $N=12$ (upper panel) and $N=16$ (lower panel) on the torus geometry at filling $\nu=2$ when tuning the interaction parameter $g_d / g_s$. The red line indicates the gap $\Delta_{\mathrm{bIQH}}$ between the lowest energy state and the first excited state. Each kink in $\Delta_{\mathrm{bIQH}}$ corresponds to a level crossing above the ground state. The blue line stands for the gap $\Delta_{{\mathrm{Pf}} \times {\mathrm{Pf}}}$ between the low energy manifold of 9 states associated to the ${\mathrm{Pf}}\times {\mathrm{Pf}}$ phase and the first excited state. In both cases, the gap is set to zero if low energy manifold that corresponds to each of these states, does not have the proper quantum numbers. The shaded area indicates the region where the lowest energy state is fully polarized (i.e. a phase separated state).}
\label{fig:phasediagram}
\end{figure}

We now turn to the phase diagram when tuning the ratio $g_d/g_s$. Figs.~\ref{fig:phasediagram}a and \ref{fig:phasediagram}b for two system sizes, $N=12$ and $N=16$. Note that the ${\mathrm{Pf}}\times {\mathrm{Pf}}$ phase occurs when $N$ is a multiple of $4$, and the NASS state occurs when $N$ is a multiple of $6$. So $N=12$ is the only case where we can have the three phases. In the $g_d/g_s=0$ limit, we observe a $9$-fold low energy manifold that corresponds to ${\mathrm{Pf}}\times {\mathrm{Pf}}$ phase. However, one cannot distinguish it for $N=16$. There is finite splitting of the $3$-fold degenerate MR ground state when this phase is realized through the two-body delta interaction. Thus when having two copies of this system in finite size, the splitting of the total $9$-fold low energy manifold can be larger than the gap of a single copy. 

The NASS state with $k=3$ and $m=2$ is a potential candidate at $\nu_t=2$. A first signature of this state should be a $10$-fold low energy manifold on the torus (the degeneracy being only approximate since we are not using the model interaction that produces this state). For $N=12$, we have never found such a signature for the full range of $g_d/g_s$ values. This should be contrasted with $\nu_t = \frac{4}{3}$ where Refs. \onlinecite{gjbl12,ueda12} found old evidence for the $k = 2$ NASS state. 

In the phase diagram, the most prominent phase is related to the one discussed for the $SU(2)$ symmetric point $g_d/g_s=1$. For both system sizes, the gap is stable over a large region of $g_d/g_s$ that include this specific point. To get a more quantitative estimate, we can compute the overlap $|\langle\Psi_{g_d/g_s=1}|\Psi_{g_d/g_s}\rangle|^2$ between the ground state any value of $g_d/g_s$ with the ground state at $g_d/g_s=1$. For $N=12$, this overlap is higher than $0.9$ for $0.80 \le g_d/g_s \le 1.25$ (i.e. centered around the peak in Fig.~\ref{fig:phasediagram}a). For $N=16$, this range becomes $0.75 \le g_d/g_s \le 1.40$ (for this system size, the Hilbert space dimension is $5.2\times 10^6$). At the peak value of gap around  $g_d/g_s \simeq 1.7$, the overlap is still $0.78$.

In summary we have studied a simple microscopic boson model that shows strong evidence for an integer quantum Hall ground state with no intrinsic topological order. This state is a 
prototypical example of a Symmetry Protected Topological insulator of bosons in two space dimensions, and is an interacting boson analog of the free fermion topological insulators. 
The existence of this interacting dominated topological insulator phase in such a simple model provides hope that other such bosons insulators in both two\cite{chencoho2011,luav2012} and three\cite{avts12} dimensions may also be realized in fairly simple models. 

Clearly a natural experimental context to seek a realization of the boson integer quantum Hall  state is in ultra-cold atoms. The delta function repulsion is realistic and natural in cold atom systems. Realizing an orbital magnetic field big enough to be in the quantum Hall regime is a long standing challenge in the field. We hope that the results of this paper provide further impetus to meet this challenge. 

\emph{Note:} During the preparation of this letter, we became aware of two related papers\cite{furukawa13,jain13}. In particular, we recover the same scaling of the gap on the torus geometry than Ref.~\cite{furukawa13}.

\emph{Acknowledgments: } 
We thank Steve Simon for enabling this collaboration and for fruitful discussions.  N.R. acknowledges Y.-L. Wu for useful discussions. N.R. was supported by NSF CAREER DMR-095242, ONR-N00014-11-1-0635, ARMY-245-6778, MURI-130-6082, ANR-12-BS04-0002-02, Packard Foundation, and Keck grant. TS was
supported by NSF DMR-1005434. TS was also partially supported by
the Simons Foundation by award number 229736. TS also thanks the hospitality of the Physics Department at Harvard University where part of this work was done.  This collaboration was initiated at the Aspen Center for Physics and was supported in part by the National Science Foundation under Grant No. PHYS-1066293.

\appendix


\begin{thebibliography}{99}

   
\bibitem{TIs} M. Z. Hasan and C. L. Kane, Rev. Mod. Phys. 82, 3045 (2010). X.-L. Qi and S.-C. Zhang, Rev. Mod. Phys. 83, 1057 (2011). M. Z. Hasan and J. E. Moore, Annu. Rev. Condens. Matter Phys. 2, 55 (2011). 

\bibitem{Fidkowski2011} L. Fidkowski and A. Kitaev, Phys. Rev. B 83, 075103 (2011).

\bibitem{Turner2011} A. M. Turner, F. Pollmann, and E. Berg,  Phys. Rev. B 83, 075102 (2011).

\bibitem{Chen2011a}  X. Chen, Z.-C. Gu, and X.-G. Wen, Phys. Rev. B 83, 035107 (2011).

\bibitem{Cirac} N. Schuch,  D. P\'erez-Garcia, and I. Cirac, Phys. Rev. B84, 165139 (2011).

\bibitem{chencoho2011} X. Chen, Z.-C. Gu, Z.-X. Liu, X.-G. Wen,  Phys. Rev. B 87, 155114 (2013).

\bibitem{LevinGu} M. Levin and Z. Gu,  Phys. Rev. B 86, 115109 (2012).

\bibitem{luav2012} Y.-M. Lu and A. Vishwanath, Phys. Rev. B 86, 125119 (2012). 

\bibitem{tsml13}T. Senthil and M. Levin, Phys. Rev. Lett. 110, 046801 (2013).

\bibitem{avts12} A. Vishwanath and T. Senthil, Phys. Rev. X 3, 011016 (2013). 

\bibitem{xuts13} C. Xu and T. Senthil, arXiv:1301.6172.

\bibitem{hmodl} C. Wang and T. Senthil, arXiv:1302.6234. 

\bibitem{maxetal13} M. Metlitski, C.L. Kane, and M.P.A. Fisher,  arXiv:1302.6535.

\bibitem{burnell13}F. J. Burnell, X. Chen, L. Fidkowski, A. Vishwanath, arXiv:1302.7072. 

\bibitem{liuxgw} Z.-X. Liu and X.-G. Wen, Phys. Rev. Lett. 110, 067205 (2013). 

\bibitem{lesik13} S.D. Geraedts and O.I. Motrunich, arXiv:1302.1436.

\bibitem{gjbl12} T. Grass, B. Juli-Daz, N. Barbern, M. Lewenstein, Phys. Rev. A 86, 021603(R) (2012).

\bibitem{ueda12} S. Furukawa, M. Ueda, Phys. Rev. A 86, 031604(R) (2012).

\bibitem{ardonne99} E. Ardonne and K. Schoutens, Phys. Rev. Lett. 82, 5096 (1999).

\bibitem{Cooper-PhysRevLett.87.120405} N. R. Cooper, N. K. Wilkin and J. M. F. Gunn, Phys. Rev. Lett. 97, 120405 (2001).

\bibitem{Regnault-PhysRevLett.91.030402} N. Regnault and Th. Jolicoeur, Phys. Rev. Lett. 91, 30402 (2003).

\bibitem{Chang-PhysRevA.72.013611} C.-C. Chang,  N. Regnault, Th. Jolicoeur and J. Jain, Phys. Rev. A 72, 13611 (2005).

\bibitem{Moore:1991p165} G. Moore and N. Read, Nucl. Phys. B 360, 362 (1991).

\bibitem{haldane91} F. D. M. Haldane, Phys. Rev. Lett. 67, 937 (1991).

\bibitem{li08} H. Li and F. D. M. Haldane, Phys. Rev. Lett. 101, 10504 (2008).

\bibitem{sterdyniak11} A. Sterdyniak, N. Regnault, B. A. Bernevig, Phys. Rev. Lett. 106, 100405 (2011).

\bibitem{furukawa13} S. Furukawa, M. Ueda, arXiv:1304.5716.

\bibitem{jain13} Y.H. Wu, J. K. Jain, arXiv:1304.7553. 

\end{thebibliography}
\end{document}